\documentclass[sigconf]{acmart} 
\AtBeginDocument{%
  }

\copyrightyear{2025}
\acmYear{2025}
\setcopyright{cc}
\setcctype{by}
\acmConference[DIS '25]{Designing Interactive Systems Conference}{July 5--9,
2025}{Funchal, Portugal}
\acmBooktitle{Designing Interactive Systems Conference (DIS '25), July 5--9,
2025, Funchal, Portugal}\acmDOI{10.1145/3715336.3735773}
\acmISBN{979-8-4007-1485-6/2025/07}

\usepackage{xcolor}
\usepackage{soul}
\usepackage{float}

\begin{document}

\title{Partnership through Play: Investigating How Long-Distance Couples Use Digital Games to Facilitate Intimacy}

\author{Nisha Devasia}
\email{ndevasia@uw.edu}
\affiliation{%
  \institution{University of Washington}
  \city{Seattle}
  \state{Washington}
  \country{USA}
}

\author{Adrian Rodriguez}
\authornote{Both authors contributed equally to this research.}
\email{aarod@uw.edu}
\author{Logan Tuttle}
\authornotemark[1]
\email{logan347@uw.edu}
\affiliation{%
  \institution{University of Washington}
  \city{Seattle}
  \state{Washington}
  \country{USA}
}
\author{Julie A. Kientz}
\email{jkientz@uw.edu}
\affiliation{%
  \institution{University of Washington}
  \city{Seattle}
  \state{Washington}
  \country{USA}
}

\renewcommand{\shortauthors}{Devasia et al.}

\begin{abstract}
  Long-distance relationships (LDRs) have become more common in the last few decades, primarily among young adults pursuing educational or employment opportunities. A common way for couples in LDRs to spend time together is by playing multiplayer video games, which are often a shared hobby and therefore a preferred joint activity. However, games are relatively understudied in the context of relational maintenance for LDRs. In this work, we used a mixed-methods approach to collect data on the experiences of 13 couples in LDRs who frequently play games together. We investigated different values around various game mechanics and modalities and found significant differences in couple play styles, and also detail how couples appropriate game mechanics to express affection to each other virtually. We also created prototypes and design implications based on couples' needs surrounding the lack of physical sensation and memorabilia storage in most popular games. 
\end{abstract}



\maketitle

\section{Introduction}
The prevalence of long-distance relationships (LDRs) between romantic couples is believed to have increased significantly over the past few decades \cite{stafford2004maintaining}, typically due to the pursuit of educational or employment opportunities \cite{Pistole_Roberts_Chapman_2010}. An estimated 25-50\% of American college students are believed to be in LDRs \cite{maguire2010distance}, and millions of married adults choose to live apart and maintain separate homes for the purpose of career advancement \cite{rhodes2002long}. Research has shown that couples in LDRs have comparable levels of relationship satisfaction to geographically close relationships (GCRs) \cite{Dargie_Blair_Goldfinger_Pukall_2015} and are equally likely to last \cite{Guldner_1996}. However, LDRs face challenges caused by a lack of co-location, such as inconsistent communication \cite{sahlstein2006making}, limited face-to-face contact \cite{dainton2001relational}, lack of intimate physical contact \cite{Neustaedter_Greenberg_2012}, and higher levels of relationship stress compared to GCRs \cite{du2016going}. 

Increased usage of social media platforms has made it possible to maintain constant lines of communication across long distances \cite{johnson2008college, hampton2017channels, Neustaedter_Greenberg_2012}. These forms of computer-mediated communication (CMC) - human communication through the use of two or more electronic devices \cite{Thorne_2008} - have become key tools for relational maintenance in LDRs, allowing couples to communicate in real time through instant messaging or face-to-face contact through video calling platforms. In particular, CMC enables couples to engage in leisure activities together, such as watching movies \cite{wanga2020social} or playing games \cite{bergstrom2009exploring}. Engaging in joint leisure activities as a couple has been shown to increase perceived relationship quality \cite{Canary_Stafford_1992, Canary_Stafford_Hause_Wallace_1993, stafford1991maintenance}.

In particular, playing games together is an increasingly common way to keep in touch with friends and romantic partners over long distances \cite{bergstrom2009exploring, Coyne_Busby_Bushman_Gentile_Ridge_Stockdale_2012, Evans_Craig_Taylor_2018}. However, digital games are relatively understudied for the maintenance of romantic relationships \cite{Evans_Craig_Taylor_2018} when compared to research on socialization with friends and family \cite{carras2017video, Choo_Karamnejad_May_2013, Ledbetter_Kuznekoff_2012, pearce2022families}, or on gaming as a coping mechanism during difficult life periods \cite{boldi2024playing, iacovides2019role, mirhadi2024playing}. These works discuss close relationships generally, and do not focus on romantic relationships. This is a notable gap in the field; particularly for young adults, romantic relationships are believed to have higher emotional valence than other close or familial relationships, and play a more significant role in emotional well-being \cite{blumenstock2022romantic}. Furthermore, while "couple technologies" have been a longstanding design space in HCI \cite{branham2013designing}, the majority of these artifacts were strictly designed for research settings \cite{Hassenzahl_Heidecker_Eckoldt_Diefenbach_Hillmann_2012}. They do not reflect more realistic communication practices used between couples, or activities that are already integrated into people's lifestyles, such as games \cite{muriel2018video}. In this work, we attempt to fill this gap by understanding how couples in LDRs use digital games as mediums of connection within their relationship. To this end, we posed the following research questions:
\begin{itemize}
    \item \textbf{RQ1}: How do different features and modalities of digital games afford the promotion of intimacy and closeness between long-distance partners?
    \item \textbf{RQ2}: What interpersonal strategies do long-distance partners use while gaming to feel closer to each other?
    \item \textbf{RQ3}: How can these insights inform the design of technologies created specifically for long-distance couples?
\end{itemize}

To explore these questions, we conducted a diary study, employed semi-structured interviews, and engaged in a design activity with 13 couples in LDRs to gain a comprehensive understanding of their experiences playing games together. The diaries and interviews explored the interplay of social dynamics between couples and their friends, benefits and drawbacks of different game modalities and mechanics, and used \citet{Hassenzahl_Heidecker_Eckoldt_Diefenbach_Hillmann_2012}'s strategies for designing technologies for long-distance connection to frame couples' gaming preferences. Through a combination of inductive and deductive analysis, we extracted key themes surrounding the affordances of cooperative vs. competitive multiplayer games, the liminality between in-person and in-game social interactions, and the appropriation of game mechanics to express affection. We developed couple archetypes to inform possible future game design interventions specifically targeted at couples in LDRs (e.g. \cite{pu2024ourhotel}), and prototyped solutions around couples' identified limitations in current game designs. We contribute a comprehensive mixed-methods, empirical understanding of how couples in LDRs are using digital games as a form of relational maintenance. 

\section{Related Work}

\subsection{Games and Relational Maintenance} Although scholars have proposed several definitions for relational maintenance (e.g. \cite{duck1999relating, dindia2003definitions, stafford1991maintenance}, all share the common theme that certain types of interactions and behaviors function to preserve ongoing relationships. Relational maintenance is used to support relationships between friends, family members, and romantic partners. However, romantic partners face particular communication challenges that platonic or familial relationships do not, and must coordinate shared commitment and life plans in addition to navigating the challenges of everyday life together \cite{shulman2013challenge}. To combat this, romantic partners engage in various relational maintenance strategies, which \citet{stafford1991maintenance}'s oft-cited conceptualization defines as: open communication, positivity, assurances, sharing tasks, and joint social interactions with friends. The advent of computer-mediated communication (CMC) brought about various networked social technologies, which are now commonly used as mediums of relational maintenance for both co-located and long-distance relationships \cite{tong2011relational}. Messaging platforms such as email, WhatsApp, and text messaging provide communication partners with "speed, relative privacy, [and] decentralization" \cite{booth2007we}, while social media and microblogging platforms such as Facebook, Twitter, and Instagram foster interactivity and connectivity with friends and strangers alike \cite{sosik2014relational, trieu2020private, quinn2018our}. Patterns of interaction remain similar between in-person and online communication: participating in the aforementioned relational maintenance strategies during CMC was shown to positively correlate with relationship satisfaction \cite{sidelinger2008couples, belus2019staying}. However, while communication is crucial to the success of any romantic relationship, it is particularly true for long-distance relationships, where in-person interactions are limited \cite{Dainton_Aylor_2002, hampton2017channels}.

Studies on relational maintenance have prioritized the investigation of romantic relationships over familial connections or friendships \cite{tong2011relational}. Conversely, gaming as a method of maintaining friendships has been relatively well studied \cite{carras2017video, Choo_Karamnejad_May_2013, Ledbetter_Kuznekoff_2012}. \citet{depping2017cooperation} found that cooperative and interdependent mechanics can promote social closeness between strangers in multiplayer games. Openness towards social connection, frequency of offline communication, and physical/social proximity have all been found to predict relational closeness in online gaming spaces \cite{Ledbetter_Kuznekoff_2012, trepte2012social}. Indeed, games are often extensions of pre-existing offline relationships (e.g. friends, family, spouses, co-workers \cite{schianoentertainment}) as opposed to isolated virtual experiences \cite{williams2006tree}, serving as "third places" \cite{steinkuehler2006everybody} where players can regularly gather to socialize and maintain their connections. However, many of these findings are general to close relationships and do not address how games can serve as relational maintenance in the face of aforementioned challenges specific to romantic relationships \cite{shulman2013challenge}. Indeed, engaging in shared leisure activities, such as games, is positively correlated to relationship satisfaction \cite{Canary_Stafford_Hause_Wallace_1993}. Furthermore, playing games together can involve all five of \citet{stafford2004maintaining}'s strategies, and we use this typology as a lens to explore how partners use games to stay connected across distances.

\subsection{Gaming During Difficult Life Experiences} Games can serve as a much needed respite from stressful circumstances \cite{iacovides2019role}. They are often used as a form of escapism, which \citet{kosa2020four} define as the avoidance of reality to allow for emotional regulation, mood management, recovery, and coping. These "four pillars" have all been explored (e.g. emotional regulation for adults \cite{bowman2012task, caro2021gaming} and adolescents \cite{gaetan2016video}; mood management \cite{rieger2015eating}; and recovery \cite{reinecke2009games, collins2014switch}), but particularly extensive research was conducted on gaming as a coping mechanism during the COVID-19 pandemic, when many people worldwide found themselves physically distanced from their loved ones \cite{boldi2024playing, mirhadi2024playing, pearce2022families, turkay2023self, barr2022playing, eden2020media}. In particular, \citet{boldi2024playing} found that players "appropriated" commercial video games in crisis situations; although most games are not explicitly designed to support resilience or psychological recovery, players imbued games with their own meanings and developed unique usage patterns to leverage them as a support mechanism in difficult circumstances. Coping has been explored in LDRs as well; \citet{maguire2010distance} explored how couples cope with problems in an exploration of college students in LDRs, using \citet{bodenmann2005dyadic}'s conceptualization of communication types: individual, dyadic, relationship-focused, and social support. The act of romantic partners gaming together could be viewed as a form of relationship-focused coping; however, none of the above papers have explored gaming as a coping mechanism for couples in particular. 

\subsection{Technology for Long-Distance Relationships} 
\subsubsection{Couple technologies}
"Couple technologies" are a longstanding design space in HCI, and researchers have created artifacts that can transmit intimate gestures through a variety of signals \cite{branham2013designing}. Some promote a sense of indirect awareness; for example, Feather and Scent, intended for when one partner was away on travel. Both artifacts can be activated remotely to make a feather float or release a scent of choice, indicating to the at-home partner that the traveling partner is thinking of them \cite{strong1996feather}. A more direct example would be inPhase \cite{tsujita2010inphase}, a system designed to provide audio signals when two users coincidentally performed the same action simultaneously. Many use light-up signals to evoke more direct sensations of awareness, such as the Virtual Intimate Object \cite{kaye2005communicating}, LumiTouch \cite{chang2001lumitouch}, or Lover's Cup \cite{chung2006lover}. Others try to simulate physical sensation, such as Hug from a Distance \cite{mueller2005hug}, inTouch \cite{brave1997intouch}, or the united-pulse rings \cite{werner2008united}. Researchers have also devised unique mediums through which couples can communicate, such as the Cube \cite{Garnæs_Grünberger_Kjeldskov_Skov_2007}, a virtual 3D cube that allows couples to compose and send encrypted symbolic messages, or Hello There, an application in which one partner can send the other an audio message from a particular geographical location, and they will only receive the message when at that particular location themselves \cite{king2007slow}. However, many of these "couple technologies" were created exclusively for research settings and are not reflective of more common methods of maintaining connection, such as video calling \cite{Neustaedter_Greenberg_2012}, texting \cite{booth2007we}, or playing games together \cite{Evans_Craig_Taylor_2018}. 

\subsubsection{Strategies for mediating relatedness}
In a scoping review of artifacts designed for relatedness, \citet{Hassenzahl_Heidecker_Eckoldt_Diefenbach_Hillmann_2012} identified strategies used to create and mediate six different facets of relatedness, which are: \textit{awareness} (creating a feeling of continuity by sharing ambient information about current activities or moods without explicit communication between partners), \textit{expressivity} (enabling expression of emotions in a wide variety of ways), \textit{physicalness} (simulating secondary effects of physical proximity or meaningful gestures), \textit{gift giving} (demonstrating caring and valuing through exchange of gifts), \textit{joint action} (allowing for carrying out an action together), and \textit{memories} (recording past activities and special moments of a relationship). Rather than defining digital games as a form of 'couple technology', however, this work frames games as an existing medium that users 'misuse' to show affection through various invented social practices. This phenomenon, in which couples in LDRs appropriate certain affordances of existing technologies to better suit their needs, has been previously documented in the literature. For example, \citet{Knobel_Hassenzahl_Lamara_Sattler_Schumann_Eckoldt_Butz_2012} document the \textit{squillo}, a social practice in Italy where a friend telephones another friend but hangs up before they can answer, indicating to the receiver that the caller is thinking of them. The telephone was not designed for this use case, but a social practice was invented around its ability to ring, and both the caller and the receiver understand the meaning conveyed by a \textit{squillo}. Similarly, \citet{Neustaedter_Greenberg_2012} investigated how long-distance couples appropriate various features of video calling platforms for intimate practices, such as kissing the webcam to simulate actually kissing their partner's forehead. Building upon this line of research, this work investigates how different features of video games are appropriated to demonstrate mutual caring. In particular, we utilize \citet{Hassenzahl_Heidecker_Eckoldt_Diefenbach_Hillmann_2012}'s six facets of connectedness as a framework for analyzing play within the context of this study. 

\subsection{Playing Video Games as a Couple}
While an exact estimate for how many couples play video games together does not exist, the percentage is likely large given the overall high participation rates among adults. In 2023, the Entertainment Software Association reported that an estimated 65\% of all Americans play video games. Of these, 74\% are adults over the age of 18 \cite{developer_2023}. Yet, the body of research on how romantic couples use games as a medium of connection is limited, and especially so when geographical separation is taken into account. A small amount of research documents the phenomenon of players meeting and becoming romantically involved through online games, with some leading to long-term and/or co-located relationships \cite{Huynh_Lim_Skoric_2013, zhang2013game, boellstorff2015coming}. Many of these couples expressed difficulties in maintaining these relationships online and over long distances \cite{Huynh_Lim_Skoric_2013}. \citet{bergstrom2009exploring} investigated how romantic couples use World of Warcraft as part of their shared leisure time; however, these couples were all co-located. \citet{Coyne_Busby_Bushman_Gentile_Ridge_Stockdale_2012} investigated how games affect conflict between couples; however, in this case, only one partner actively played games, and geographical co-location was not specified. Indeed, much of the work on couples gaming together has been focused on how gaming affects relationships in which one partner (typically male) plays games frequently and the other (typically female) does not \cite{Coyne_Busby_Bushman_Gentile_Ridge_Stockdale_2012, ahlstrom2012me, cole2007social}. \citet{schianoentertainment} suggested that many female players begin playing games due to male spouses or romantic partners. \citet{yee2006demographics} found significant differences in player motivations between gender, with male players prioritizing \textit{achievement} and \textit{manipulation}, and female players prioritizing \textit{relationships}. More recently, \citet{poeller2018let}'s Motive Disposition Theory suggests that implicit player motivations may have a greater effect on in-game social behavior than explicit motivations. Researchers have investigated how to address such asymmetries in desired outcomes and styles of play through game design (e.g. \cite{harris2016leveraging, gerling2014effects}). These asymmetries extend to relationship satisfaction;  \citet{ahlstrom2012me} suggested that relationship satisfaction was negatively impacted when only one partner engaged in gaming, but that playing games together had a positive effect. Indeed, participating in leisure activities together has been shown to increase relationship satisfaction \cite{sidelinger2008couples}. To this end, \citet{Evans_Craig_Taylor_2018} used the Relational Maintenance Strategies Measure \cite{Stafford_Dainton_Haas_2000} to deductively analyze couples’ survey responses regarding their experiences playing League of Legends together; however, only 57.6\% of couples reported being in long-distance relationships. A further limitation of the current body of research is that many of these works are focused on specific games, and are limited primarily to the study of Massively Multiplayer Online Role Playing Games (MMORPGs). This work intends to generalize across broad features of games from many genres that couples play together, rather than investigate dynamics within a particular game. Additionally, we seek to address the difficulties expressed by couples who maintain the majority of their relationship through games due to geographical separation \cite{Huynh_Lim_Skoric_2013}. 

 \section{Methods}

To understand the experiences of long-distance couples using digital games as a medium of connection, we conducted a diary study \cite{rieman1993diary} and semi-structured interviews with 13 couples spanning a wide range of geographical separation and relationship length. 

\subsection{Participants and Recruitment}
We recruited using a screener survey, which was disseminated through gaming servers at our university and posted on social media sites such as Reddit and Twitter. This survey asked basic questions about couple demographics, what games they tended to play together and how often, and how they felt playing games affected their relationship. Inclusion criteria specified that participants must currently live more than 50 miles away from each other in separate residences, must be over 18 years of age, and must play games with their partners at least once a week. Of the 75 initial responses to the screening survey, 15 couples were contacted to participate in the full study. In choosing these couples, we sought a variety of age ranges and gender identities, types of games played, and a range of time zone differences. One couple did not complete any study activities past the initial onboarding, and one couple separated during the study, resulting in 13 couples who completed the entire study (see Table \ref{tab:participants}). These 13 couples had an average age of 25.2, with the majority of participants in their early to mid 20s, which is within the typical age range for long-distance relationships \cite{maguire2010distance}. They had been together for an average of 2.34 years, ranging from 2 months to as long as 8 years. Five couples had no time zone differences but lived geographically separate (e.g., Seattle and San Francisco), while other couples reported time differences of up to 19 hours (e.g., the Philippines and the USA). While six couples had at least one member outside of North America, only one couple was completely international. The majority see each other in person about once a month, but 2 couples were ‘never mets’, which means that they met and began their relationship online but had not met in person at the time the study was conducted. This phenomenon is documented in studies such as \citet{Huynh_Lim_Skoric_2013}, \citet{boellstorff2015coming}, and \citet{zhang2013game}. The 26 individuals who completed the study were each compensated with a \$40 USD gift card. 
\begin{table*}
  \caption{Overview of participants who completed all portions of the study}
  \label{tab:participants}
  \begin{tabular}{cccclll}
    \toprule
    PID &Age & Gender & Location & Relationship length & In-person meetings & Joint favorite game (Platform) \\
    \midrule
    C1A & 25 & M & USA & 6 months & Yearly & Cine2Nerdle (Browser) \\
    C1B & 23 & NB & Australia &  &  &  \\
    \midrule
    C2A & 23 & M & Seattle & 3 years & 3x a year & Splatoon 3 (Nintendo Switch) \\
    C2B & 23 & F & San Francisco & & & \\
    \midrule
    C3A & 25 & F & Seattle & 2.5 years & Monthly & Scribblio (Mobile) \\
    C3B & 22 & M & San Francisco & & & \\
    \midrule
    C4A & 22 & NB & USA & 3 years 4 months & Yearly & Splatoon 3 (Nintendo Switch)\\
    C4B & 22 & F & United Kingdom & & & \\
    \midrule
    C5A & 24 & F & New York & 2 years & Monthly & We Were Here (Desktop) \\
    C5B & 24 & M & Boston & & & \\
    \midrule
    C6A & 27 & F & Philippines & 1.5 years & 2x a year & Genshin Impact (Desktop) \\
    C6B & 27 & M & USA & & & \\
    \midrule
    C7A & 24 & F & Canada & 3 years & 3x a year & Final Fantasy XIV (Desktop) \\
    C7B & 27 & M & USA & & & \\ 
    \midrule
    C8A & 31 & F & USA (2 hour difference) & 2 months & Every few months & Stardew Valley (Desktop) \\
    C8B & 27 & NB &  &  &  & \\
    \midrule
    C9A & 24 & NB & USA & 2 years & Only met once & Minecraft (Desktop) \\ 
    C9B & 22 & F & Canada & & & \\
    \midrule
    C10A & 29 & F & Turkey (500 km apart) & 8 years & Every several months & Heroes of the Storm (Desktop) \\
    C10B & 36 & M &  & & & \\
    \midrule
    C11A & 31 & M & USA & 5 months & Never mets & Apex Legends (Desktop) \\
    C11B & 26 & F & United Kingdom & & & \\
    \midrule
    C12A & 26 & F & USA (100 miles apart) & 4 years & Monthly & Stardew Valley (Desktop) \\
    C12B & 27 & M & & & & \\
    \midrule
    C13A & 22 & M & Germany & 2 months & Never mets & Out of Space (Desktop) \\
    C13B & 27 & NB & USA & & & \\
    \bottomrule
  \end{tabular}
\end{table*}

\subsection{Data Collection}
The data for this study was collected from February - May 2024. Participants from the initially selected 15 couples all attended a 15 minute onboarding session with the lead researcher, in which the goals of the study and the two main parts of the study (the diary study and the semi-structured interview) were explained. All participants completed a consent form approved by our institution's Institutional Review Board (IRB) prior to the onboarding session.

\subsubsection{Diary Study} Participants were asked to keep a diary in which they logged a minimum of five entries each after playing a game with their partner. In the onboarding session, they were instructed to fill out the diary (provided through Google Forms) immediately after they finished playing a joint gaming session. The goal of this was to have concrete, in-the-moment examples from couples' play sessions to probe later on in the interviews, as well as to encourage couples to reflect on their own play. Each diary entry asked the following questions:
\begin{itemize}
    \item Today, my partner and I played the following game: [Open-response]
    \item I would describe this game in the following ways: [Options: Cooperative, Competitive, Goal oriented, Requires a lot of grinding, Requires strategizing, Casual, Creative, Other]
    \item How long did the game session last for? [Open-response]
    \item Describe some of the gameplay you engaged in today. [Open-response]
    \item While playing in today's game session, some of the emotions I felt were: [Open-response]
    \item On a scale of 1-5, playing today made me feel: [1: very disconnected from my partner, 5: very close to my partner]
    \item Describe the most memorable part of the gameplay today and why it was memorable. [Open-response]
\end{itemize}

In addition, diary entries asked participants to identify which of \citet{Hassenzahl_Heidecker_Eckoldt_Diefenbach_Hillmann_2012}'s six facets of connectedness, if any, they felt applied to their game session. We adapted and simplified definitions from the original paper to position the game as the technology promoting the six facets of connectedness. They were provided to the participants as follows:
\begin{itemize}
    \item \textit{Awareness}: The game enabled me to feel a sense of continuity and presence with my partner without doing anything specific.
    \item \textit{Expressivity}: The game enabled me to express my feeling and emotions in a wide variety of ways.
    \item \textit{Physicalness}: The game allowed for a feeling of physical intimacy.
    \item \textit{Gift giving}: The game gave me the ability to demonstrate caring and valuing my partner through reciprocal exchange.
    \item \textit{Joint action}: The game allowed my partner and I to carry an action out together.
    \item \textit{Memories}: The game allowed us to keep records of our activities and special in-game moments.
\end{itemize}

\subsubsection{Semi-Structured Interview} \label{semistructure}
After completing their diary entries, participants engaged in a final semi-structured interview with the lead researcher, lasting around 60-75 minutes. Interviews were conducted with both members of the couple simultaneously to glean additional detail on their diary entries and overall experience gaming together. The semi-structured interview protocol was tailored to each couple’s diary entries, and questions were therefore variable across couple; however, common questions included:
\begin{itemize}
    \item What do you like/dislike about playing cooperative/competitive games with each other?
    \item What sorts of topics do you discuss while gaming together?
    \item How does playing with friends affect the dynamic between you and your partner?
    \item Which of Hassenzahl’s six facets of connectedness (presented to each couple) are the most important to you in the games you play and why? 
    \item Which facets would you like to see implemented more in games and why?
\end{itemize}

\subsubsection{Mechanics, Dynamics, Aesthetics Activity} \label{mdaactivity}

\begin{figure*}
  \centering
  \includegraphics[width=\textwidth]{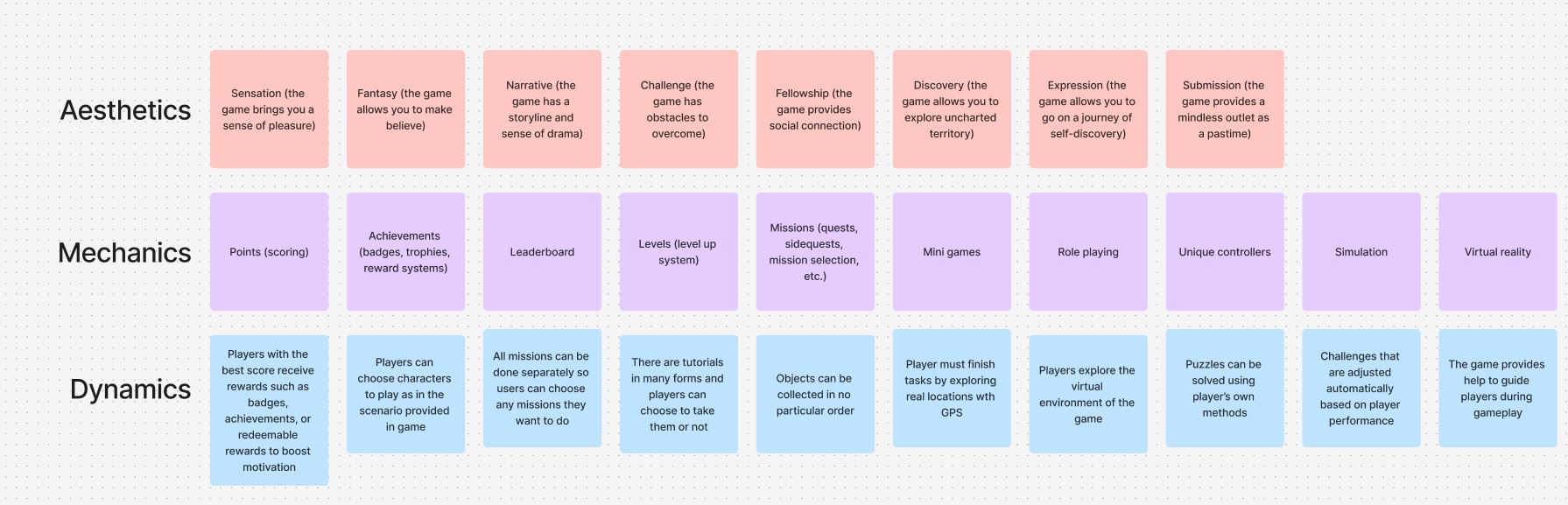}
  \caption{A snapshot of the initial board given to participants for the MDA activity, listing examples of each component (the full list of mechanics and dynamics are not shown in this image).}
  \label{mdainitial}
  \Description{}
\end{figure*}

\begin{figure*}
  \centering
  \includegraphics[width=\textwidth]{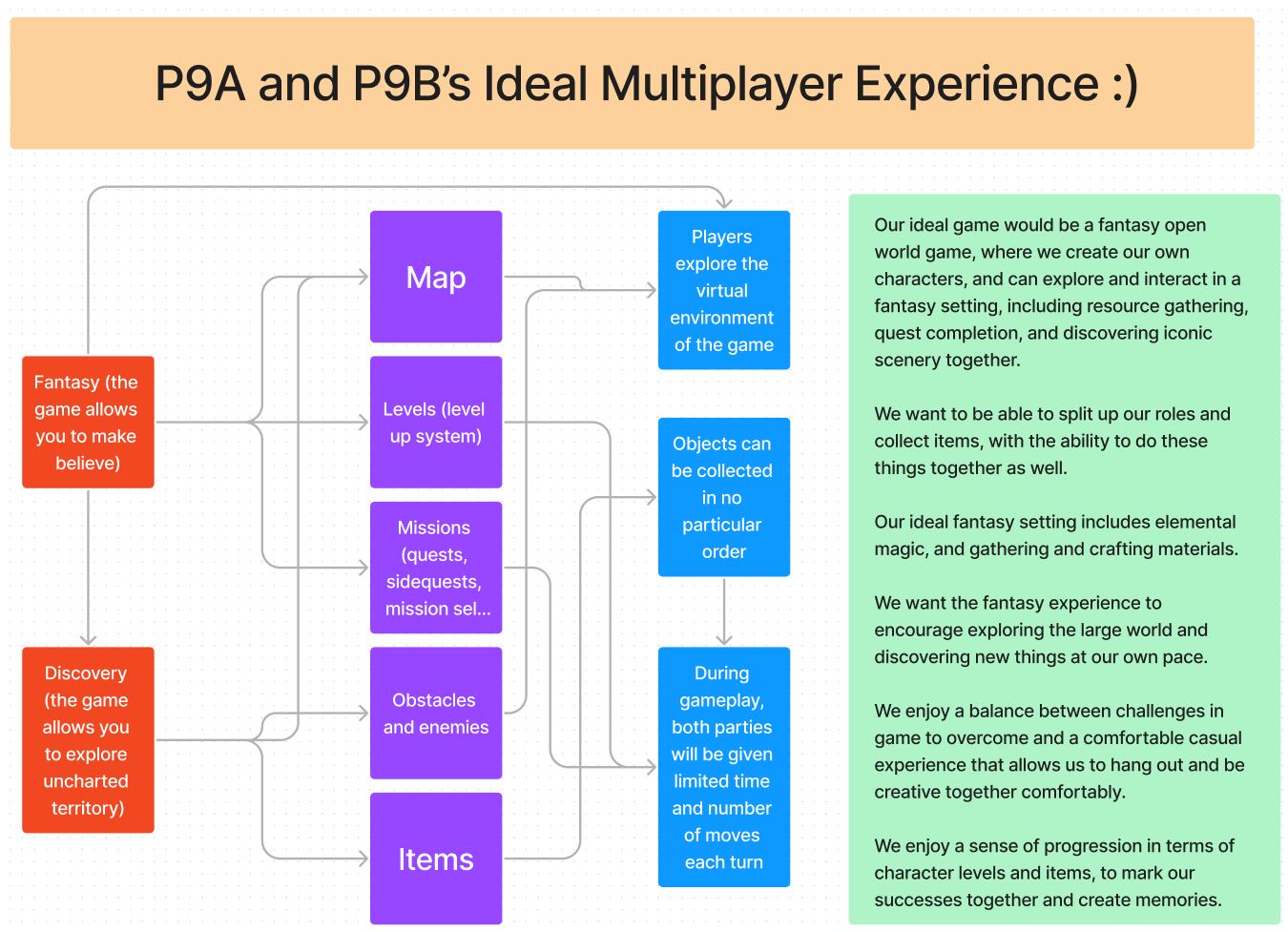}
  \caption{An example of the output of a couple’s MDA activity. The ideal aesthetics are in red (Fantasy, Discovery), mechanics are in purple (Map, Levels, Missions, Obstacles/enemies, Items), and dynamics are in blue (Players explore virtual environment of game, Objects can be collected in no particular order, Turn-based gameplay).}
  \label{mdaexample}
  \Description{}
\end{figure*}

At the end of the semi-structured interview, couples jointly engaged in an activity using the MDA (Mechanics, Dynamics, Aesthetics) framework. First presented by \citet{Hunicke_LeBlanc_Zubek_2004} at the Game Developers Conference in the early 2000s, the MDA intends to bridge the gap between game design, criticism, and research. We chose to adapt this commonly used framework as an activity for non-experts by discretizing each component of the MDA into individual elements and presenting them to participants as a Figma board of sticky notes (see Figure \ref{mdaexample} for what participants were initially presented with). We derived a starting set of elements from \citet{Kusuma_Wigati_Utomo_Putera_Suryapranata_2018}, and invited participants to add their own elements as they saw fit. As suggested in \citet{buttfield2016better}'s implementation of the MDA framework in an informal environment: "because the player perceives the game through the lens of its aesthetics, the selection of which aesthetics should be evoked by the game makes for a useful starting point for the game developer". Therefore, we had participants - acting as both players and designers within the scope of the activity - begin by identifying their ideal aesthetics, then mechanics, and finally dynamics. The lead researcher began the activity by presenting the Figma board to each couple and giving them 30 minutes to identify elements of an ideal game to play together, and did not participate in this activity other than answering clarifying questions about different elements. See Figure \ref{mdaexample} for an example of a couple's output from this activity. 

\subsection{Analysis}

We conducted the semi-structured interviews over Zoom and used Zoom's in-built audio transcription or Rev.ai to generate transcripts. The first author, along with a research group consisting of five undergraduate and masters students, performed a qualitative analysis of the interview transcripts. For the first four transcripts reviewed, the group followed the steps of \citet{Braun_Clarke_2006}'s thematic analysis: Each member inductively open-coded the same transcript independently, and then brought their codes to the research group meeting. Researchers then made sticky notes for each individual code, and then performed an affinity diagramming exercise (also known as the KJ method \cite{scupin1997kj}) to develop themes. After four iterations, the research team reviewed and agreed upon a set of preliminary themes using a consensus model. The first author then used these preliminary themes to independently code the remainder of the transcripts. No new themes were added, but the preliminary themes were subdivided and elaborated upon with data from the additional transcripts.

We analyzed the diary entries with Microsoft Excel and wrote Python and R scripts for further statistical analysis. These results are presented as a descriptive overview in Section \ref{diaries}.

Through our analysis of the diary entries and interview transcripts, we found that the 13 couples in our study fell into three distinct categories based on how they described themselves playing games together. These archetypes are presented in Section \ref{archetype}. After developing definitions of each archetype, the first and second authors, as well as three members of the qualitative research group, independently assigned each couple to an archetype, based on their diary entries and interviews. Intercoder reliability among five raters using Fleiss’ kappa was calculated as 0.67, indicating substantial agreement from Landis and Koch's heuristic \cite{Landis_Koch_1977}. 

In order to determine if couple's preferred playstyles aligned with certain game features, we used the archetype classifications to analyze the outputs of each couple's MDA activity, described in Section \ref{mdaactivity}, to determine frequently referenced mechanics, dynamics, and aesthetics for each archetype. This analysis is presented in Section \ref{implications}.

\section{Findings}

\subsection{Descriptive Overview of Diary Entries} \label{diaries}
The 13 couples in the study contributed a total of 76 diary entries. Of these, we excluded 5 due to only one partner recording an entry about a specific game session, resulting in a total of 71 entries included for analysis. The open questions on the diary were used to inform the specific questions asked to the couple in the semi-structured interview following completion of the diary entries (see Section \ref{semistructure}), and as such, are not part of the analysis below.  

\begin{figure*}
  \centering
  \includegraphics[width=\textwidth]{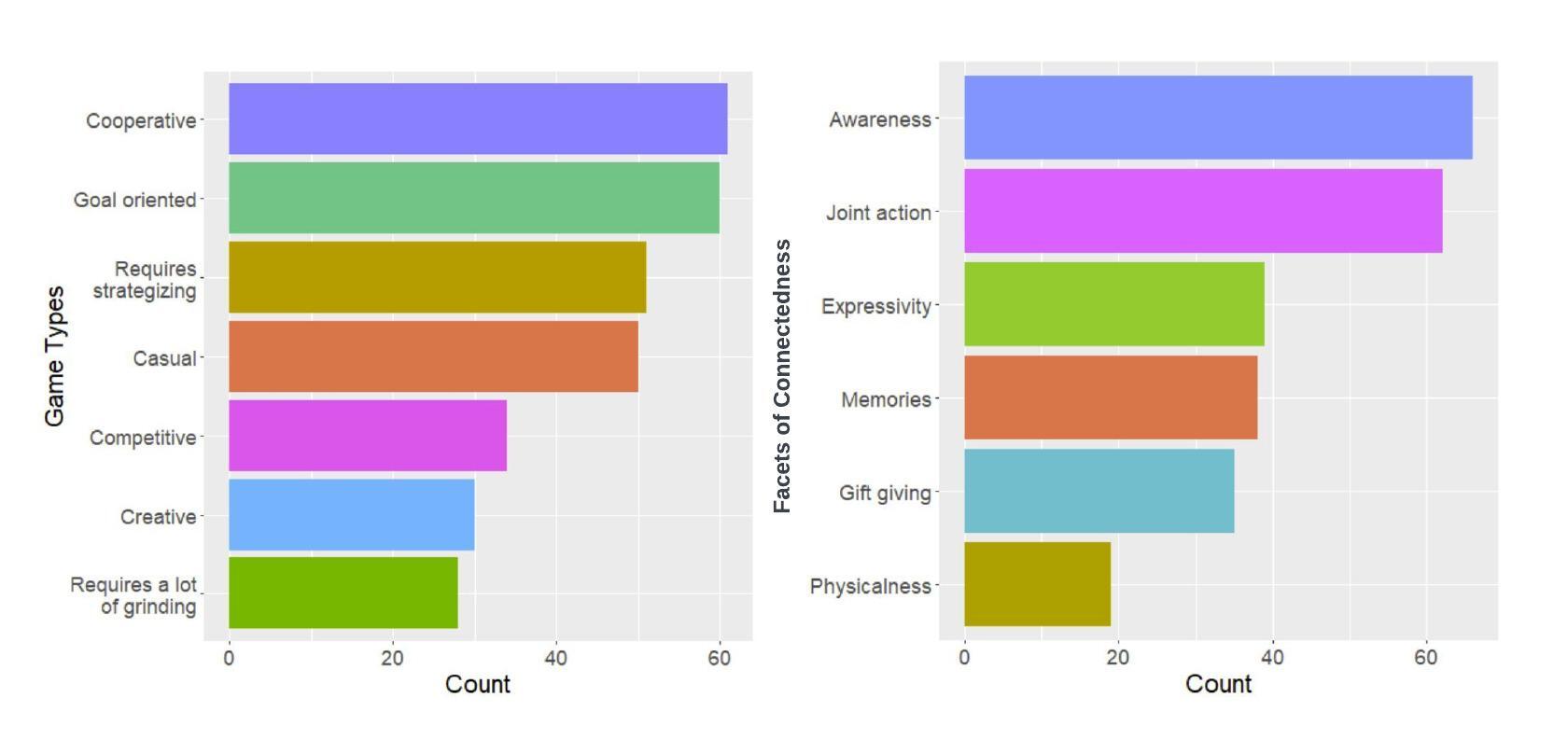}
  \caption{How often game types (left) and facets of connectedness (right) were reported in the diary entries.}
  \label{fig:4.1histo}
  \Description{}
\end{figure*}

\subsubsection{Games} Participants listed a total of 32 unique games in the entries. By far the most frequently mentioned game, listed in 18.3\% of entries, was \textit{Stardew Valley}. \textit{Stardew Valley} is designed to be a relaxing, cooperative farming simulator considered to be the archetypical "cozy game", characterized by feminist and inclusive design \cite{waszkiewicz2020towards}. Other frequently played games included \textit{Splatoon 3} (7\%), a cooperative player-versus-environment (PvE) shooting game, and \textit{It Takes Two} (5.6\%), a cooperative action-adventure game following the story of a couple on the verge of divorce.
\subsubsection{Connection} We asked participants to report in each session entries how connected they felt to their partner on a Likert scale from 1-5, where 1 was extremely unconnected and 5 was extremely connected. Average connection reported between partners was 4.401/5. Using Pearson's coefficient, connection reported by each partner in a couple was found to be moderately highly correlated, ($r(69) = 0.607, p < 0.001$). Together, these results imply that partners felt similarly high levels of connection when gaming together. 

\subsubsection{Game Types} The diary form asked participants to choose from the following game modalities to describe the game played in their session (Figure \ref{fig:4.1histo}). The majority of game sessions involved cooperative (85.9\%) or goal-oriented (84.5\%) games. 

\subsubsection{Facets of Connectedness} The diary form also asked participants to identify which of Hassenzahl's six facets of connectedness they felt applied to the game played in their session (Figure \ref{fig:4.1histo}). Almost all sessions involved games that participants believed to promote awareness (92.9\%) or joint action (87.3\%). 

\subsubsection{Time Played}
Play sessions lasted for an average of 2.096 hours. The Kruskal-Wallis H test indicated that there was no significant difference in playtime between the game types, but a significant difference in playtime between facets of connectedness, $\chi^2(5) = 11.23, p = .047$ (Table \ref{tab:desc}). A possible explanation for this is that games participants listed as promoting expressivity (e.g. \textit{Pictionary} clones, \textit{Scrabble} clones) were turn-based competitive games played for a few rounds per gaming session, whereas games listed as promoting physicalness and memories (e.g. \textit{Stardew Valley}, \textit{Minecraft}) are more conducive to being played for several hours in a single session.

\begin{table}
  \caption{Average playtime reported for entries tagged with each facet of connectedness.}
  \label{tab:desc}
  \centering
  \begin{tabular}{|c|c|}
    \hline
    \textbf{Strategy} & \textbf{Average time played (hours)} \\
    \hline
    Awareness & 2.159 \\
    \hline
    Gift Giving & 2.588 \\
    \hline
    Expressivity & 2.091 \\
    \hline
    Joint Action & 2.265 \\
    \hline
    Memories & 2.661 \\
    \hline
    Physicalness & 3.131 \\
    \hline
  \end{tabular}
\end{table}

\subsection{Why couples value different game modalities} Games that allow for two or more players are normally characterized by dichotomies of cooperative/competitive and interdependent/independent mechanics \cite{depping2017cooperation}. Cooperative games primarily allow (or require) players to complete objectives together within the same game world (e.g., crafting towns and cities together in the building game \textit{Minecraft}, maintaining a farm together in popular cozy game \textit{Stardew Valley}, or working together to complete food orders in a high octane kitchen in \textit{Overcooked}). Competitive games are normally characterized by having players compete against each other, whether 1-on-1 (e.g., chess), a team vs. another team (e.g., Multiplayer Online Battle Arenas like \textit{League of Legends}), or multiple people vying for the same objective (e.g., \textit{Jackbox Games}). In our interviews, we investigated how couples viewed cooperative vs. competitive mechanics and how they used them for relational maintenance.

\subsubsection{The value of cooperative games} \label{coopvalue}
\begin{quote}
    "\textit{Being able to work together towards the same goal and celebrate our victories together is really great.}" - C2B
\end{quote}
As long-distance partners are unable to engage in \textit{shared tasks} \cite{stafford1991maintenance} such as chores or errands in person, cooperative games with interdependent mechanics served as a replacement medium for couples to work towards and complete goals together (C3, C5, C7, C8, C9). As C9B explained, \textit{“Working towards a goal, especially in games, invites that feeling of fulfillment and accomplishment which can…lead into that feeling of closeness. Because it's like, hey, we did this, and we did it together. It’s that feeling of being able to accomplish things together [and see] our progress together.”} Completing these goals together was also accompanied by a sense of shared accomplishment, which was cited as an important reason for enjoying working on joint goals, as \textit{“the shared celebration in meeting the goal together brings [the couple] closer”} (C4B). The presence of a goal was also preferable to some couples, who desired objectives \textit{“instead of aimlessly walking around”} (C7B). 

To accomplish these goals, couples often found themselves communicating to divide labor and complete tasks. Division of labor often occurred along lines of what each partner preferred to do; for example, C10A stated that she prefers to grind and collect items in every game she played with her partner and that this works well for her because she does not enjoy the combat elements that C10B prefers. It was also crucial for in-game tasks that are tedious to complete on one’s own. C2 discussed the collection of resources in farming simulator \textit{Stardew Valley}, particularly the harvesting of wood from their forest: \textit{“Coming in and being like, okay, this is what we want to do, let’s split up the work to make this happen…And that was a lot nicer to together than individual because it’s just so much time.”} Indeed, certain game mechanics require partners to rely on each other directly to complete a task. C13 discussed their playthrough of \textit{It Takes Two}, in which, in certain levels, one partner must shoot nails into the wall in a precise configuration. This allows the other partner to use those nails as platforms that can be climbed up to unlock the next area. This task cannot be completed with just one player, and C13A expressed his appreciation for the game forcing him to rely on his partner: \textit{“You actually have this feeling [that you are] a couple when you need to work together...you need to really rely on each other. I like this aspect that you are a team.”}

More than half of the couples (C3, C5, C6, C7, C8, C11, C12) also discussed how the communication required to accomplish these shared goals also helped their communication skills overall. Often, this was because achieving the objective was difficult, like completing a virtual escape room together (e.g., C5 playing \textit{Escape Simulator}), determining what they could have done differently when they failed a mission (e.g., C7 playing \textit{Astroneer}), or needing to put together several parts of the story to understand the full picture (e.g., C8 playing \textit{Bokura}). However, couples also put their communication to use in different ways; for example, overcoming frustration together. As C3A recalled, \textit{“We'll play a game called Bread and Fred, where it's like two penguins trying to climb up and it's sometimes really difficult. And sometimes we just have to keep trying over and over and over again to get over a hurdle. I think in those moments where we do, we are able to not be frustrated at each other, but [instead] we're able to try to achieve a goal together and we achieve it.”} C3B also highlighted the importance of positive encouragement in the face of frustrating situations: \textit{“I try to encourage her and…she would encourage me. There's a lot of communication involved in that game, but [that] wouldn't be [true] if we were just silent and trying to figure out the game ourselves. You have to really talk a lot. I think [cooperative] games encourage us to be more cheerful, playful and talkative.”} C12A echoed this sentiment and discussed an instance where she felt guilty for accidentally spending a large portion of their shared bank account in \textit{Stardew Valley}, but was reassured by C12B that it was okay: \textit{“Especially in a low stakes moment, like a game…it's still nice to get the reassurance, versus in a very high stakes real life moment, where it's stressful [and] it's hard to take that [reassurance] in sometimes. But in the game, I can see that [reassurance].”}

\subsubsection{The value of competitive games}
\begin{quote}
    “\textit{I think it’s healthy in relationships to have a little competition every once in a while.}” - C10A
\end{quote}
Many couples (C1, C2, C3, C4, C6, C8, C11) stated that the primary reason they played games with competitive elements was for the \textit{positivity} and \textit{assurances} \cite{stafford1991maintenance} that the low stakes competitive environments provided. C4B described the humor she found in losing to her partner: \textit{“I take competitive games pretty lightheartedly, so it's not as if I get upset or anything. I think it's funny when I die. I think it's funny when he dies. I think it's funny when we trade and we both kill each other. It's a nice playful feeling to have a one-up over him or jokingly having beef with each other.”} Even partners who identified as being highly competitive or bad losers stated that they were able to let go of these aspects of their personality when it came to gaming with their significant other. For example, C1B, despite stating, \textit{“I’m not a great loser with things in real life. I’m the kind of person who has to win everything”}, cited instances in their diary entries where they allowed their partner to win the game for the humor of the situation. C3A reported a similar sentiment: \textit{“I’m a competitive person. Maybe it’s petty, but I like to beat him sometimes. But then we get really silly over it and even if we name call each other, it’s all lighthearted, not anything serious.”} The reassuring nature of not getting angry or frustrated at each other was discussed within the context of competitive games as well. C4B, who identifies as a woman, expressed that: \textit{“It's also reaffirming as well when you have that kind of dynamic in a competitive environment, because let's be real, men are scary when it comes to competitive games. But having a time where you play competitive games where neither of you actually get upset at each other for losing [is] reaffirming in the dynamic that you share.”} 

To maintain the positivity that competitive games could engender, most couples maintained that they only chose games where partners were relatively balanced in skill level. For example, C3A and C3B, who primarily play competitive games individually as well as together, discussed only jointly playing games that they were both equally familiar or equally unfamiliar with. C5A, despite never playing competitive games either individually or with C5B, expressed a very similar sentiment: \textit{“If we were to play competitive games..it would have to be something that we are like both equally adept [and] there would have to be like like some chance involved. So that way, it [would] even it out. Say [C5B] was incredibly great at something. And I was really bad, or I just had no skill in doing it. I feel like I would not like doing that.” }

\subsection{Liminality between real life and gameplay}
\begin{quote}
    \textit{"Doing a limited time event together is...something we set the time apart specifically to do. In this way I feel like it's similar to real life, where going to a limited time event together to enjoy something has merits for a relationship building experience, just by having that joint in game time doing something we can't always take for granted."} - C9A
\end{quote}

In the absence of in person time together, couples have found ways to treat certain aspects of the games they play together as parallels to real life shared activities. Some couples (C2, C6, C9) likened their game time to going on a date; for example, when asked as to the value C6 derives from menial in-game tasks such as raids versus the value of open-world exploration, C6B used the analogy, \textit{“It’s like doing chores [together] versus going on a date.”} C9B furthered this analogy, comparing limited-time events in games, which require players to log in and play the game during certain time periods, to visiting an exhibit at a museum: \textit{“In person, it would be like, ‘Hey, let’s go to this exhibit together, and walk through that [together].’ Whereas digitally, it’s: ‘Hey, there’s this thing that’s up, and we can explore it together.’ And I think that’s a really nice way of interacting.”} C9A further elaborated, \textit{“The limited aspect makes [us] have to set a scheduled time to do it. I think that in itself does add an aspect to it where it does feel more like we would on a date.” }
	
Games also provide couples with \textit{“constant opportunit[ies] to come up with new silly things”} (C9A), primarily inside jokes and topics of conversation that they discuss outside of their time spent playing together. Several couples (C2, C4, C9, C12, C13) referenced inside jokes that they had with each other that bled beyond the boundaries of the game and into their \textit{“everyday vocabulary”} (C9B). These references surfaced during other activities that the couples engaged in together, such as watching a TV show together (C13), during an in-person activity (C9), or with other friends. 
 
Indeed, almost every couple engaged in \textit{joint social interactions with friends} \cite{stafford1991maintenance}. Since many couples have friends across different countries and time zones, the games and the digital spaces created to play them (e.g., Discord servers) are key ways to stay connected socially. In fact, some couples met and began their relationship in these spaces; for example, \textit{“The [Discord] server was how I met [C9B], because of that [Minecraft YouTuber] fan group. It’s been a really nice way to spend time with her, and also other friends with similar interests.”} (C9A). Social dynamics in games reflected experiences that partners may have with groups of friends in real life. Positive aspects included spontaneity introduced by the addition of more friends to a play session (C6, C7, C9), or using the game sessions a way \textit{“to spend time with people (especially post-COVID)”} (C2A). Some couples reflected on the negative aspects of playing with their friends as well; C2B reflected on how playing with friends, as opposed to playing 1:1 with C2A, \textit{“takes up a little more of [her] social battery”}, while C7 discussed how playing with friends who are worse at the game can detract from their enjoyment.

Games were also a way for certain participants to meet their partner’s friends and integrate into existing friend groups, a crucial rite of passage in building a romantic relationship. C6B recounted his experience as the partner being introduced to C6A’s friend group: \textit{“It was really nice to get me introduced to some of her friend group, because I didn’t really know these people until we spent three hours running around the open world together, and that was a nice bonding experience. Since I am not as social of a person as [C6A], sometimes I need a little bit of an incentive to go out and meet new people.”} C11A recalled his opposite experience introducing C11B to his gaming friend group: \textit{“[C11B] fits in pretty well…I’ve had an ex before who just did not fit into my friend group, so I would never play with both of them at the same time. So this is really nice.”} 

Couples also stated that the goals of their play sessions 1-on-1 with their partners, as opposed to playing together with friends, were different altogether. As C2A summarized, \textit{“I think the focus [when playing with friends] is split across multiple people, versus very intentionally spending time with one person.”} Indeed, many couples reflected on how playing with friends was mainly a way to \textit{“put a little bit of time and energy into that friendship”} (C2A), while playing 1-on-1 with a partner was treated as \textit{“couple time”} (C13A). These sessions were set aside specifically for focusing on their partner and their relationship. Individual sessions also gave partners a chance to be more verbally affectionate and \textit{“speak more freely”} (C11B), rather than facing \textit{“teas[ing]...about being disgusting lightheartedly”} (C9B). 

\subsection{Expressing affection through game mechanics}
\begin{quote}
    \textit{"Even as they were busy fighting the aliens...[C13B] still found time to splash water on me [to save me], and I was grateful and felt taken care of for this."} - C13A
\end{quote}

While the individual gaming sessions themselves gave couples a safe space to express their verbal affection for their partner, many couples (C1, C2, C3, C4, C9, C10, C12, C13) discussed ways in which they additionally used game mechanics to express more physical feelings of fondness, something which a lack of co-location makes impossible for long-distance couples. C10A recounted her appreciation for C10B secretly crafting a graduation hat for her in \textit{Palworld} and presenting it to her after she graduated from her masters degree. Similarly, C13A expressed how he felt \textit{“taken care of”} in the instances that C13B would notice his low health bar before he did himself or when C13B came to his rescue in battles despite being overwhelmed with enemies himself. C2B also discussed the \textit{“protectiveness”} he felt towards C2A: \textit{“I prioritize rescuing [C2A]...this [protectiveness] is something that you can do in a physical space, but not as much virtually.”} Even relatively minor physical actions in the game world, e.g., waiting for a partner’s avatar to catch up while exploring an area (C6, C13), or helping a partner collect resources for their in-game project (C9), helped participants feel closer to their partners. Partners would also \textit{“let [their partner] win”} (C1B), or play games that the other partner specifically enjoyed because it brought them joy to watch their partner enjoy themselves (C4). 

Couples also appropriated game mechanics in ways that did not necessarily align with design intentions. Notable examples included C3 playing \textit{Spellcast}, a game similar to Scrabble, in which they defined the winner as the person who could create the most words related to their inside jokes. When asked to explain this practice, C3A said, \textit{“We know we’re making each other laugh…that internally is our highest level that we can achieve.”} Defining the heuristic for success as making the other partner laugh was a common practice upon couples to reinforce shared intimacy. In the context of a creative drawing game, in which players must guess what the other person is trying to draw, C1B recalled, \textit{“[In the game] I genuinely cannot draw anything. Whereas [C1A] is very good at drawing. So the [drawings that C1A makes] that I can guess are because he can draw well, but the [drawings that C1B makes] that he can guess are because he knows me well and I just find that really funny.”} Similarly, C13A stated that when playing drawing games with friends, he could always guess which drawings were C13B’s because \textit{“[C13B]’s sense of humor, or sometimes [they would draw] inside jokes, like how I love penguins.”} 

\subsection{Framing play with Hassenzahl et al.'s facets of connectedness} \label{framing}
\begin{quote}
    \textit{“I guess that's one of the things about playing games, is that you know they're doing this to spend time with you. You’re doing something together that does not require you to be like, oh, she's still there, or is she still interested. Or something that would just quite make you question whether or not they value the current thing that you both are currently doing.”} - C7B
\end{quote}

Analysis of the diary entries showed that awareness and joint action were the most frequently reported of \citet{Hassenzahl_Heidecker_Eckoldt_Diefenbach_Hillmann_2012}’s facets of connectedness. This was displayed in the qualitative results as well: when asked which facets they found the most important to their play, 10 couples said awareness and 9 said joint action. Participants provided several reasons for the high value placed on awareness, primarily surrounding the \textit{“sense of the position or presence of another person”} (C3A), even without necessarily actively talking over voice chat. Participants also gave several examples of how game mechanics promoted awareness, such as \textit{“passing each other in the game while we’re on our way to different tasks”} (C12B), or \textit{“doing stupid stuff in the game with our characters, like jumping up and down…that creates that sense of presence”} (C3B). As C13A summarized: \textit{“I think it’s important for the game to give you a sense that you are playing with your partner live.”} 

While mechanics promoting awareness in games afford a more passive sense of presence, participants also expressed appreciation for mechanics that promote active presence through joint action. C5A expressed that: \textit{“I like joint action [because] in order for me to do a task, he has to be here…cause you’re doing the same thing at the same time with each other.”} C5A provided \textit{It Takes Two} as an example of a game with such mechanics (see Section \ref{coopvalue}). C9B presented an explicit comparison between joint action and awareness: \textit{“Carrying out a joint action almost invites that feeling of closeness more because you’re working towards a goal together rather than just hanging out and vibing.” }

When asked about which facets they would like to see implemented more in games, several participants mentioned that they rarely encountered strategies promoting memories; for example, C8A said that \textit{“[they] were surprised that memories didn’t come up very often in the games [they] played”}, and C5B expressed that \textit{“nothing stuck out to [them] as...a game that definitely incorporates that”}. A potential reason for this may be that popular gaming platforms execute strategies for memories poorly. For example, C2A mentioned that \textit{“the [Nintendo] Switch system for sending screenshots to your personal devices really should be better.”} C9B expressed a similar problem with the Steam platform: \textit{“A lot of people will take screenshots and then they just won’t know where to find them. That happens to me a lot. I’d taken screenshots in [Don’t Starve], and they were buried with the Steam folder, and then in the game apps folder, and then in the game folder, and then there’s a screenshot folder. It’s just so hard to find. If more games had a more accessible way of looking at it, I think that would definitely help.”} Due to the lack of functional memory saving, several participants expressed that they save special memories manually; for example, C1 and C2 discussed how they screenshot or take pictures of the game screen with their phones to save memories, C4B described \textit{“a Discord channel dedicated to quotes and screenshots and special moments”}, and C9B described their elaborate system for memory saving: \textit{“I have all my screenshots sorted into folders for every individual playthrough, with different co-op modes, with individually modded playthroughs…occasionally when I open it up, I’ll be like, here’s a screenshot of when we built this together.”} Indeed, there was an expressed need for better implementations of strategies for memories. As C11A stated, \textit{“I think I would like to see more of memories…it would be nice to have cute little screenshots of our characters or places or something that we made together that we can refer back to.” }

Physicalness was the least reported facet in the diary entries. When participants discussed it, it was primarily in the context of being able to interact with their partner’s avatar (e.g., \textit{“In games like Minecraft, where you’re actually playing as a character interacting in the world…the character is like their own person.”} - C9B), or examples of mechanics that translate to physical affection (e.g. \textit{“It’s nice to be able to hug [C5B] [in Stardew Valley] as much as I want…it takes a few seconds and interrupts whatever you’re doing. In real life I’m pretty physically affectionate and so I feel like that aspect of our real relationship is translating.”} - C5A). However, in \citet{Hassenzahl_Heidecker_Eckoldt_Diefenbach_Hillmann_2012}’s framework, strategies promoting physicalness revolve around firsthand physical sensation, primarily mediated through haptic experiences. While some console controllers and mobile devices have recently implemented haptic feedback that corresponds with in-game events, these are primarily for single player games. In addition, the majority of desktop games lack the requisite hardware for haptics. These are gaps that are open for further exploration in design and research contexts.

\subsection{Developing couple play style archetypes} \label{archetype}

Our analysis yielded three play style archetypes: 1) couples who strictly preferred playing cooperative games that prioritize joint action, 2) couples who play a balanced mix of competitive and cooperative games, and 3) couples that prefer competitive games for the afforded expressivity. These were not archetypes that the research team had preconceived and we did not describe or prompt participants to identify with any of these profiles. While these archetypes and associated facets of connectedness could be applied to any romantic couple gaming together, we found the archetypes to be particularly helpful in brainstorming design prototypes for long-distance couples, presented in Section \ref{recs}.

\subsubsection{Archetype 1: Strongly prefers co-op games, strongly prioritizes joint action} 
A plurality of couples (46\%) strongly prefer games with cooperative elements over those with competitive elements, exhibiting a preference for gameplay that promotes collaboration and mutual trust-building. Notably, couples in this archetype tended to explicitly avoid playing competitive games against their partner. This was often due to a perceived imbalance in gaming skill between partners, e.g. C10A: \textit{"I don’t really perform well in competitive games. And [C10B] likes competitive games and he is really good at it. That’s why I don’t feel like I’m contributing enough."} C5A echoed this sentiment, stating: \textit{"I'm really bad at competitive stuff. And I'm not one to do things that I'm very bad at. I'm not adventurous in that way."} These sentiments align with previous findings regarding skill differentials in social gaming contexts \cite{gerling2014effects}. Couples in this archetype tended to prefer games such as \textit{Stardew Valley}, characterized by its casual, cooperative, and repetitive nature. They tended to highly value awareness and joint action and placed a strong emphasis on teamwork, shared goals and accomplishments, and mutual support. These attributes align strongly with \citet{yee2006demographics}'s \textit{Relationship} motivation, where "willingness
to form meaningful relationships that are supportive in
nature" is prioritized. Additionally, their preferred elements from the MDA activity, such as 'having a storyline and background story' and 'fantasy' (see Figure \ref{fig:mdaresults}) indicated a motivation for \citet{yee2006demographics}'s \textit{Immersion} ("enjoy creating avatars with histories
that extend and tie in with the stories and lore of the
world").

\subsubsection{Archetype 2: Balancing co-op and competitive, prioritizes awareness}

Couples who fall under this archetype (38\%) are characterized by their preference for cooperative games, although they are not exclusively committed to them, as they balance their preference by occasionally engaging in competitive gaming. Although similar to Archetype 1 in terms of valued facets of connectedness, namely awareness and joint action, couples in this archetype tended to engage in more competitive gameplay together, whether competing on a team in games such as \textit{Splatoon 3} or \textit{League of Legends}, or casually competing against each other in games such as \textit{Jackbox}. Like Archetype 1, couples in this archetype seem to be motivated by the \textit{Relationship} aspects of play, but their preferred MDA elements, namely 'missions', 'mini games', and 'object collection' (Figure \ref{fig:mdaresults}) indicate that they also value \textit{Achievement} ("accomplishing goals and accumulation of items that confer power") \cite{yee2006demographics}.

\subsubsection{Archetype 3: Strongly prefers competitive, strongly prioritizes expressivity} A minority of couples (16\%) displayed a preference for 1v1 gameplay, exemplified by casually competitive trivia games such as \textit{Skribbl.io} (a \textit{Scrabble} clone) or drawing games similar to \textit{Pictionary}. Couples in this archetype stated that they enjoy casual competition and suggest that competitive gaming offers more opportunities for playful jokes, teasing, and banter, which increase overall engagement and enjoyment. These aspects partially align with \citet{yee2006demographics}'s \textit{Manipulation} ("enjoy deceiving, scamming, taunting, and dominating"), albeit with lighthearted intentions. Their preferred MDA elements, namely 'points', 'increasing difficulty', and 'receiving rewards' (Figure \ref{fig:mdaresults}) also showed a motivation for \textit{Achievement}. Although both couples in this archetype played cooperative games, they were the only couples to explicitly mention perceived drawbacks of playing cooperative games together. For example, C3A said that she believed that the \textit{"communication aspect [of cooperative games] can be emotionally exhausting after a while, especially if you're [upset]."} C3B added that he likes being \textit{"bound by his own skill level"} and values the autonomy of most competitive games. Indeed, both couples in this archetype consistently listed expressivity as a key design feature in the games they play together.  

\subsubsection{Connecting Archetypes with MDA Activity} \label{implications}

\begin{figure*}
  \centering
  \includegraphics[width=\textwidth]{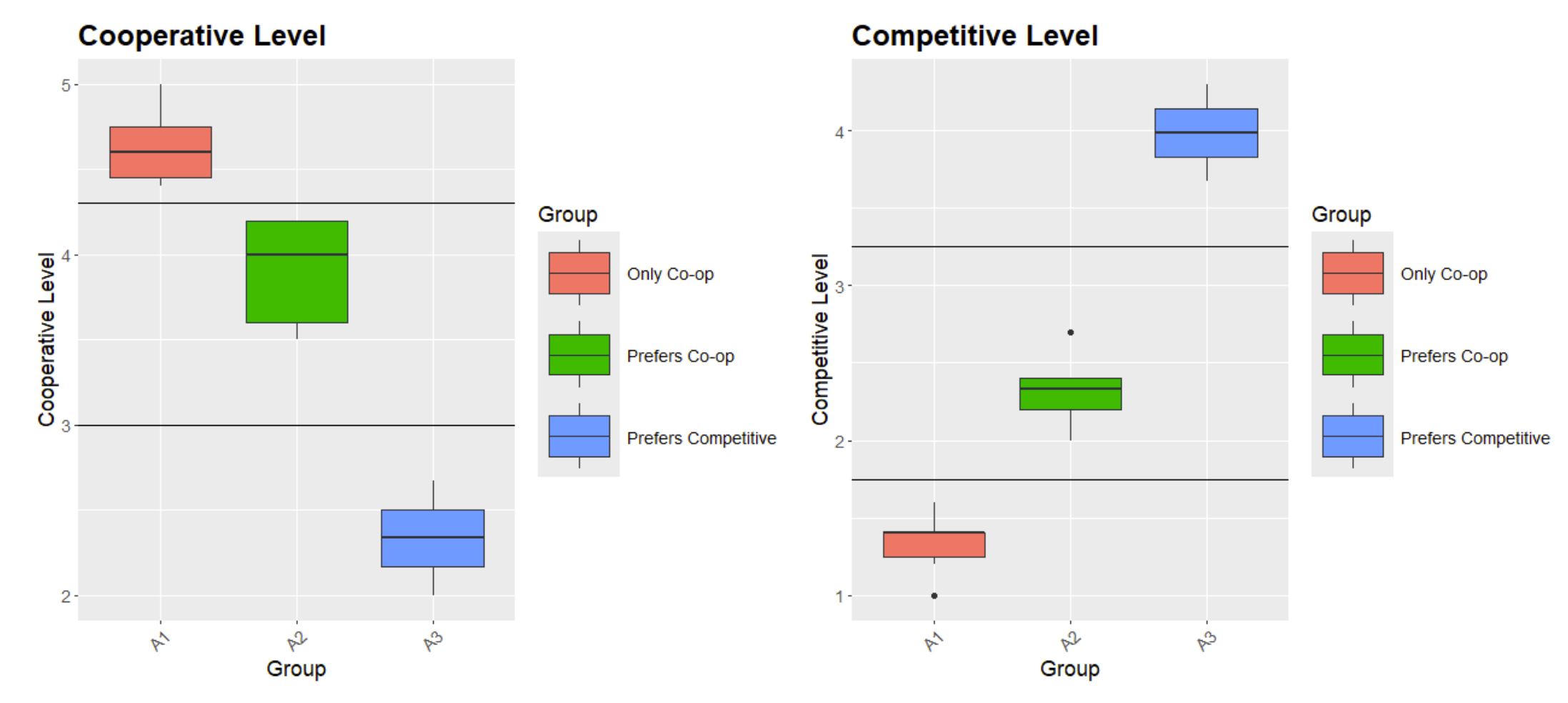}
  \caption{The five raters independently gave each couple a Likert score ranging from 1-5 on two axes: cooperative and competitive. Averaging these ratings for each archetype showed clear divisions between each archetype on these axes.}
  \label{fig:boxplots}
  \Description{}
\end{figure*}

\begin{figure}
  \centering
  \includegraphics[width=\linewidth]{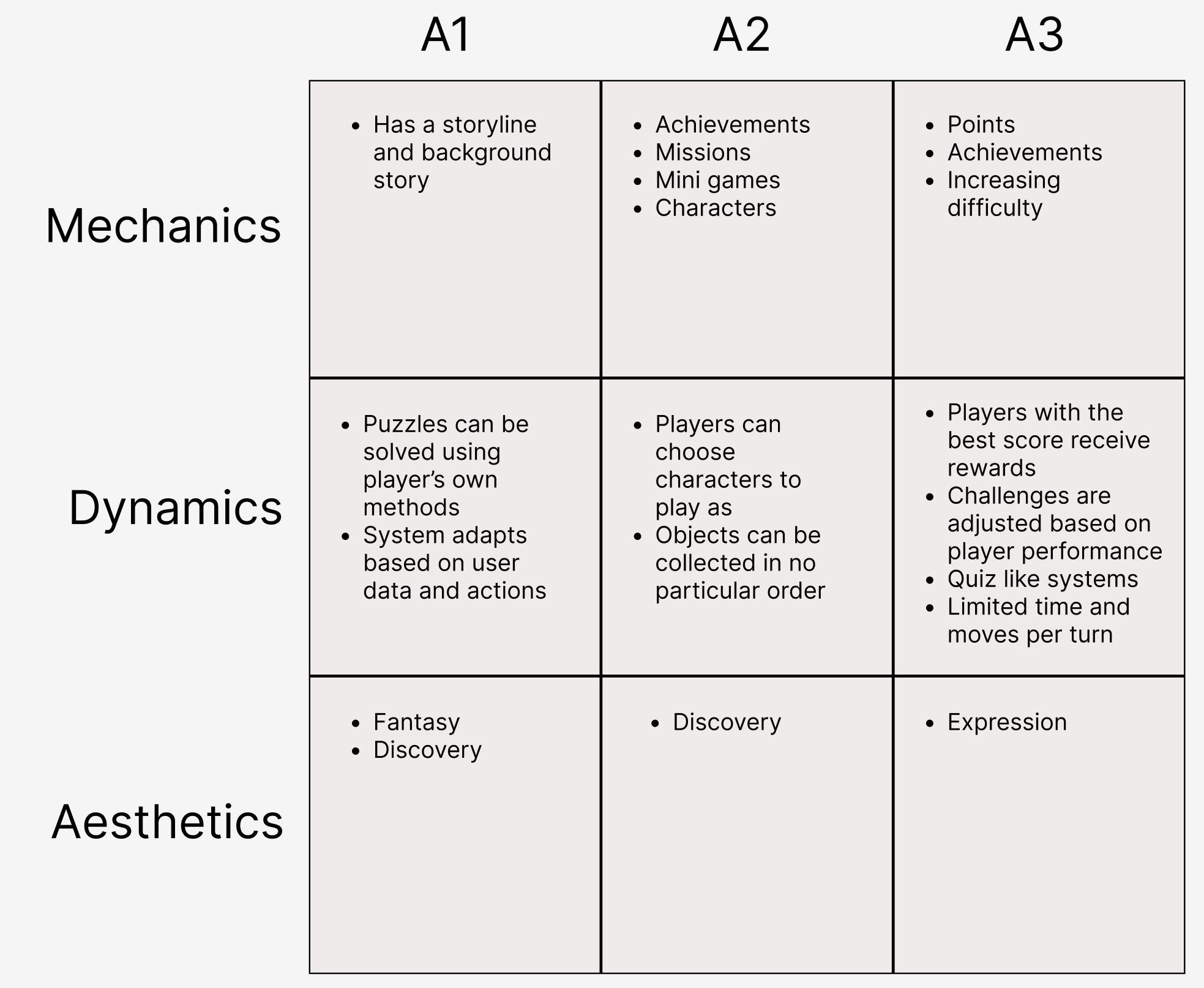}
  \caption{Commonly used mechanics, dynamics, and aesthetics referenced across archetypes in the MDA activity.}
  \label{fig:mdaresults}
  \Description{}
\end{figure}

\begin{table*}
\centering
\begin{tabular}{|c|c|p{10cm}|}
\hline
\textbf{Couple ID} & \textbf{Archetype} & \textbf{Description of ideal game to play together} \\ \hline
C1 & A3 & "Like It Takes Two with more 1v1 aspects and the ability to screw the other player over" \\ \hline
C2 & A2 & "Age of Empires with horses to make it less war simulator and more whimsical; Animal Crossing with killing" \\ \hline
C3 & A3 & "SIMS if you could roleplay as seals competing to be the biggest" \\ \hline
C4 & A2 & "Multiplayer Breath of the Wild" \\ \hline
C5 & A1 & "It Takes Two combined with the open world and character creation aspects of Baldur's Gate, with puzzles like those in The Witness" \\ \hline
C6 & A2 & "Genshin Impact if it was an MMO" \\ \hline
C7 & A1 & "Final Fantasy XIV but more dynamically adjusted so that casual players can still complete a lot of content" \\ \hline
C8 & A2 & "Combination of the mechanics of Wobbly Life and the character and story depth of Bokura" \\ \hline
C9 & A1 & "Combination of the open world aspects of Genshin Impact and the gathering mechanics and limited time activities of Stardew Valley" \\ \hline
C10 & A1 & "Red Dead Redemption with sandbox functionality; Skyrim with more memorable characters and better narrative" \\ \hline
C11 & A1 & "Borderlands if it took place in the world of Apex Legends" \\ \hline
C12 & A1 & "Coffeeshop simulator where one person takes care of all the management and inventory and the other interacts with the NPCs" \\ \hline
C13 & A2 & "It Takes Two but with collectibles or trophies for doing certain tasks (similar to Stardew Valley mechanics)" \\ \hline
\end{tabular}
\caption{Couples' descriptions of their ideal game to play together, based on their preferred mechanics, dynamics, and aesthetics listed in the MDA activity.}
\label{tab:ideal_games}
\end{table*}

\begin{figure*}
  \centering
  \includegraphics[width=0.8\textwidth]{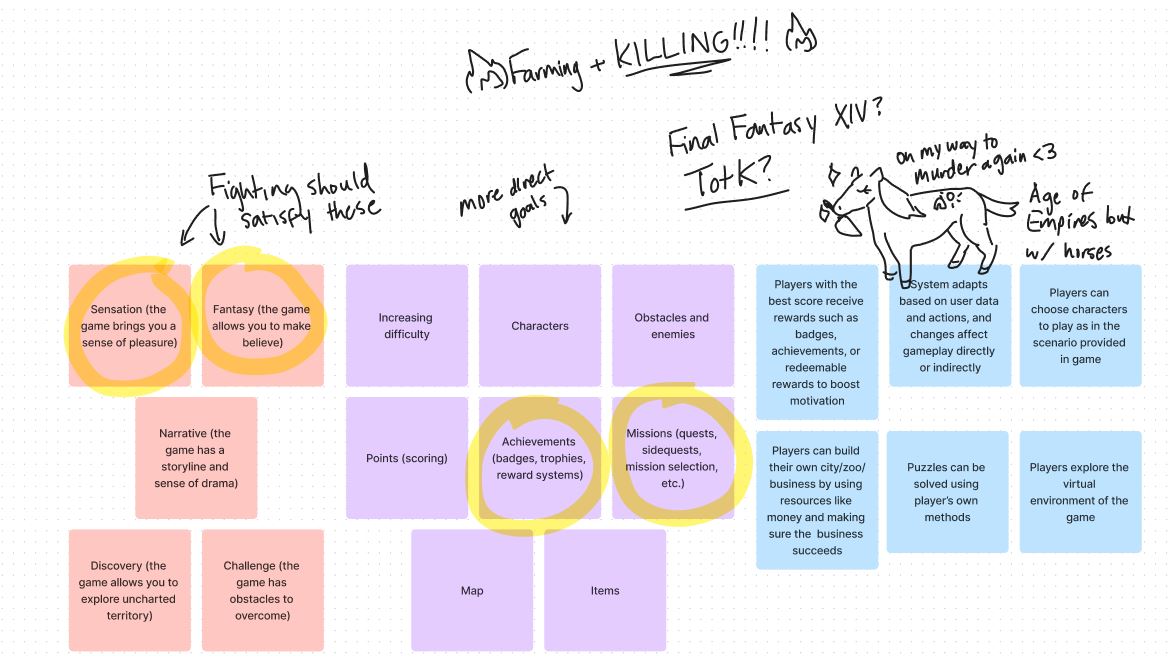}
  \caption{Design artifact from C2 in Archetype 2 describing their ideal game.}
  \label{fig:ageofempires}
  \Description{}
\end{figure*}

Couples' preferences in the MDA activity aligned with the developed archetype descriptions (see Figure \ref{fig:mdaresults}), with A3 primarily choosing features of competitive games (e.g., points, turn taking, rewards for high scores). When asked to describe their preferred games, they expressed a desire for 1v1 gameplay, e.g., C1B expressing that they \textit{"would only want [the co-designed game] to be against each other"}, C1A valuing \textit{"the ability to mess with the other player"}, or C3 describing their ideal game as \textit{"the SIMS but with seals competing to be the biggest."} A1 prioritized exploration and story-based games, e.g., \textit{"Borderlands (a cooperative story game with exploring and level progression) with more relatable characters, or if Apex Legends (a battle royal multiplayer game) had a story mode"} (C11). Couples in A2 chose a mix of both cooperative and competitive elements, e.g., C2 describing their ideal game as \textit{"Age of Empires (a real time strategy game where players gather resources to fight opposing factions) but with horses"} (see Figure \ref{fig:ageofempires}). A full list of couples' descriptions of their ideal games to play together is presented in Table \ref{tab:ideal_games}.  

\section{Discussion}
Our research explored how couples in LDRs value different modalities and mechanics in digital games, and how they use games as a medium through which they can simulate real life activities and express affection to each other. We also developed couple archetypes to represent different values that couples have around their own play.

In the following sections, we discuss implications for HCI researchers and game designers who are seeking to create new technologies for couples in long-distance relationships. As the number of LDRs are believed to be on the rise, this is a crucial period to consider how we, as researchers and designers, can support this lived experience through game design and the creation of game-adjacent applications.

\subsection{Games as relational maintenance} 

Our findings build upon \citet{Evans_Craig_Taylor_2018}'s investigation of couples' exhibited maintenance behaviors in League of Legends, which showed that couples reported high relational quality due to their time spent gaming together. We extend those findings by investigating long-distance couples specifically, and how they engage in relational maintenance in a wider range of games. Indeed, relationship maintenance behaviors are key to the success of long-distance relationships \cite{dainton2001relational}. Broadly, the couples in our study used games as a shared leisure activity, an important relational maintenance strategy \cite{Canary_Stafford_Hause_Wallace_1993}. Moreover, we found that participants' in-game behaviors mapped to \citet{stafford1991maintenance}'s typology of relational maintenance strategies. In the \textit{positivity} dimension, participants generally "act[ed] cheerful and positive" and "[tried] to be romantic, fun, and interesting" with their partners during play. More specifically, participants were "patient and forgiving of [their partner]" (such as C12A's anecdote about accidentally spending a large portion of C12's shared Stardew Valley money), and "attempt[ed] to make [their] interactions very enjoyable" (such as C1B allowing C1A to win even though it is against their competitive nature, or C3 defining making their partner laugh as the heuristic of success). In the \textit{assurances} dimension, participants primarily appropriated game mechanics to "show his/her/their love to [their partner]", such as through C10B's crafting of a presenet for C10A, C13B risking their in-game safety for C13A, or C9 helping each other with individual in-game projects. In the \textit{network} dimension, participants "focused on common friends and affiliations" through joint play sessions with friends, and simultaneously displayed to their partners that they "like[d] to spend time with the same friends". In the \textit{tasks} dimension, participants showed their partners that they could "help equally with tasks that need to be done" and "share in joint responsibilities" through collaboration in cooperative games, such as the division of labor discussed by C2 and C10. Finally, in the \textit{openness} dimension, games served as a space for participants to "tell [each other] how they/he/she feels about the relationship" and to effectively enact "what he/she/they need/want from the relationship" through one on one playtime, as discussed by C2, C11, and C13. 

It is important to note the particular role of games for these long-distance couples. The high degree of liminality to real life that games afford is crucial: it serves as a medium through which to engage in a much wider range of relational maintenance behaviors than otherwise available to them. While \textit{positivity}, \textit{assurances}, and \textit{openness} can be expressed through common means of computer-mediated communication such as messaging or calling, it is difficult to engage in \textit{network} or \textit{task} behaviors. However, our participants showed how they are filling these gaps through their joint play. Conversely, while co-located couples may also use games as a form of relational maintenance, they have a wider range of options available to them through which to fulfill maintenance behaviors; i.e. in-person dates, social hangouts, or sharing a physical space. Games are one of the only mediums through which long-distance couples can engage in these maintenance behaviors, and therefore merit further exploration in the context of maintaining - and thriving in - long-distance relationships. 

\subsection{Appropriation of games and their mechanics to express affection}

\citet{Huynh_Lim_Skoric_2013} discuss how "intimate relationships are mediated heavily by the symbolic representations in the game itself". We extend those findings with practices similar to those described by \citet{Knobel_Hassenzahl_Lamara_Sattler_Schumann_Eckoldt_Butz_2012} and \citet{Neustaedter_Greenberg_2012} regarding the ‘appropriation’ or ‘misuse’ of technology to express intimacy and affection, in ways likely not intended by the designers. \citet{Hassenzahl_Heidecker_Eckoldt_Diefenbach_Hillmann_2012} describe such practices as: “a product of people’s inventiveness to fulfill their needs even in the face of 'inappropriate' technological solutions”. Similarly, \citet{boldi2024playing} describe how players "adopted usage patterns and ascribed meanings to video games that were not inscribed in their original designs" during difficult circumstances. We witnessed this inscription of meaning through the adjustments that couples made to suit their relationship dynamics, or to embed personal meaning and shared histories into their gaming experiences. Descriptions of joint gameplay also highlighted participants’ deep understanding and recognition of their partner’s likes and preferences. These inventive practices also underscore the importance of mutual recognition and support in relationships in instances where participants would prioritize their partner’s happiness and satisfaction over their own within the game. 

It is notable that there are very few popular games designed explicitly for couples, let alone couples in long-distance relationships. Of the games mentioned in our sample, \textit{It Takes Two} likely comes the closest to meeting these specifications, as it is specifically designed to be played with two people and the two player characters are a couple. Long-distance technologies in general have been thoroughly explored in HCI research, yet few of these findings have been translated to game design. We can draw upon our findings as well as learnings from HCI research to suggest modalities and mechanics that may promote connection between couples in games. For example, the theme of shared inside jokes or secret languages has been explored previously and was found to create a strong feeling of relatedness \cite{cheal1987showing}. Game designers could implement an object akin to \cite{Garnæs_Grünberger_Kjeldskov_Skov_2007}'s \textit{Cube} within games as a collectible, which allows the couple to both engage in the joint action of discovering the object and enabling them to develop their own "inside" language. Multiplayer features akin to \cite{king2007slow}'s \textit{Hello There} exist (e.g., custom scavenger hunts in \textit{Minecraft}, player generated messages hidden around the world in \textit{Elden Ring}), but the mechanics could be tuned such that only a specific player would be able to activate a specific object or message, thus allowing partners to hide gifts for each other. In a related vein, just as C10B crafted a graduation hat for C10A to surprise her with, mechanics could be added to generate crafting recipes based on player preferences or real-life events such as birthdays or other celebrations. Drawing upon the joint action strategy, game designers could also consider adding more event-based or limited-time features to simulate the feeling of a "date" as described by C9, and including special collectibles that represent their shared participation in the event (e.g., matching avatar outfits). 

While these examples represent \citet{Hassenzahl_Heidecker_Eckoldt_Diefenbach_Hillmann_2012}'s facets of connectedness as applied to a subset of our findings, designers could use this framework to further surface potential ways in which players appropriate game mechanics in novel ways and use those findings to design game modalities specifically targeted at long-distance couples.

\subsection{Design Recommendations} \label{recs}

\subsubsection{Investigating on a per archetype basis}

While \citet{depping2017cooperation} suggest that cooperative and inderdependent mechanics in games are more effective in promoting social closeness, the behaviors and preferences expressed by Archetype 3 (competitive, prioritizes expressivity) suggest that more investigation may be warranted regarding how competitive mechanics factor into positive behaviors in close dyadic relationships. Indeed, our findings show that long-distance couples are not monolithic in the ways they play games together, as evidenced by the significant differences in play time and preferred game types between our developed couple archetypes. These differences would also have an effect on designing for/with couples with strong preferences for different modalities of games. This not only includes different game mechanics (see Section \ref{implications} for examples), but extends to interfaces or products designed for different couple archetypes as well. Values around play styles, implicit and explicit motivations \cite{yee2006demographics, poeller2018let}, preferred game mechanics, and tool design for couples in LDRs could be explored through participatory design activities, building on our findings from the MDA activity and archetype classifications. 

\subsubsection{Addressing gap around "memories" facet: Prototyping third party application for memory storing} \label{memoriessection}
\begin{figure}
  \centering
  \includegraphics[width=\linewidth]{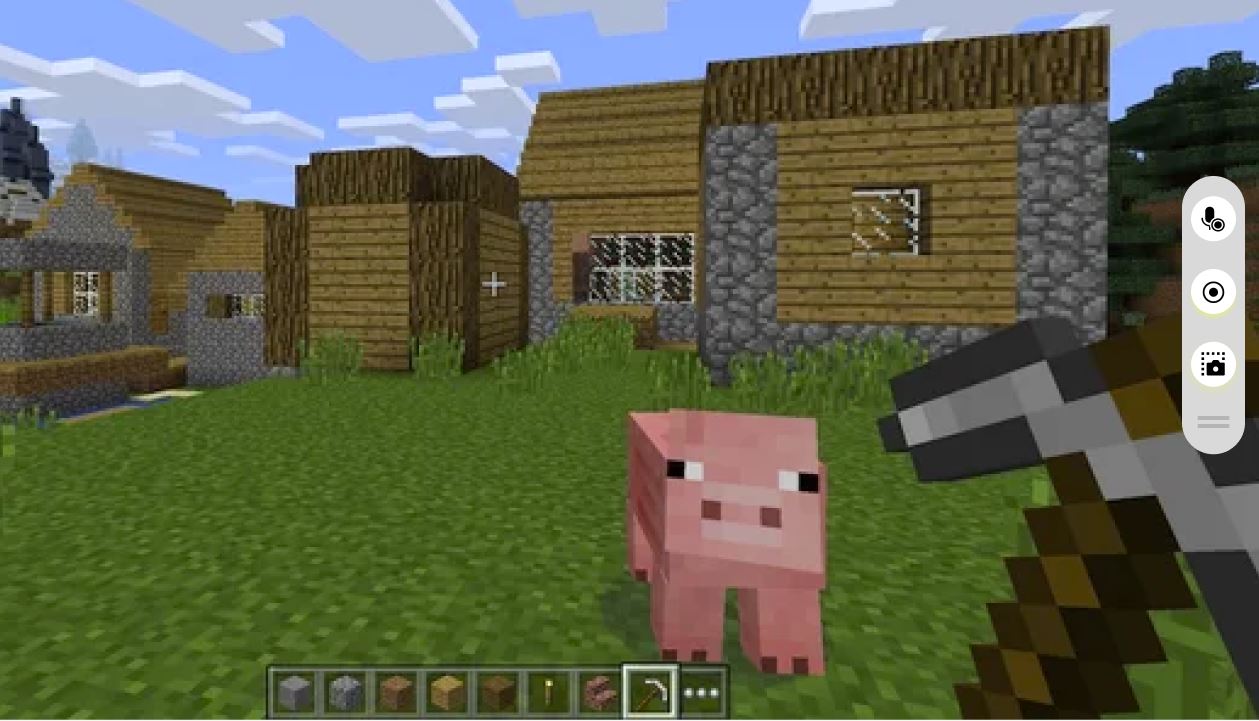}
  \caption{The desktop overlay, shown on the right side of the screen, allows users to easily take screenshots or screen recording from their gameplay. These are then saved into the diary application.}
  \label{fig:overlay}
  \Description{}
\end{figure}

\begin{figure}
  \centering
  \includegraphics[width=\linewidth]{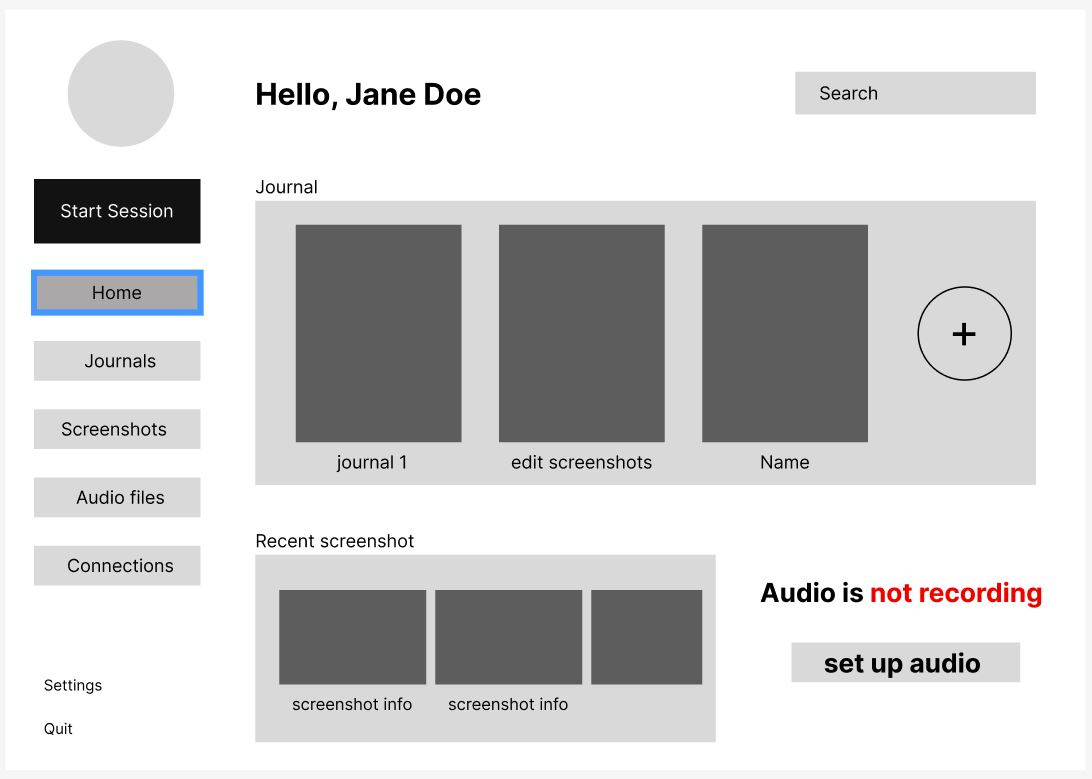}
  \caption{An initial low-fidelity prototype of the "digital diary" application home screen, developed by our research team.}
  \label{fig:homepage}
  \Description{}
\end{figure}

To address the user needs expressed regarding better memory saving functionality for digital games (see Section \ref{framing}), our research group prototyped a desktop overlay (Figure \ref{fig:overlay}) connected to an application in which couples can store screenshots and recordings of their joint play sessions (Figure \ref{fig:homepage}). The goal of such an application would address the lack of standardization between game or platform photo functionalities and provide a single location where users could access shared photos and recordings. Drawing inspiration from other cloud photo services that allow users to save and curate their images, such as Apple iPhotos or Google Photos, the proposed "digital diary" application would allow couples to organize memories into game-specific journals and allow for minor image editing functionality such as adding text boxes, stickers, or GIFs. 

\subsubsection{Addressing gap around "physicalness" facet: Participatory design work for peripheral design} \label{haptics} Although researchers have devised numerous haptic devices to simulate physical intimacy over long distances \cite{mueller2005hug, brave1997intouch, werner2008united}, few of these designs have relied on grounded observations or participatory design. Our findings show that players value “awareness” more than any other of \citet{Hassenzahl_Heidecker_Eckoldt_Diefenbach_Hillmann_2012}'s six facets of connectedness, and simultaneously engage with “physicalness” less than any other facet. Given that researchers have developed numerous hardware peripherals to promote awareness, the low incidence of physicalness in our findings suggests that desktop game makers are underusing haptic technologies. Moreover, accounts of the irreplaceability of ‘physical protection’ (C2B) and the ‘emotional exhaustion’ (C3A) of constant verbal communication during cooperative play reveal opportunities for further research into interpersonal haptics. The diversity of gameplay forms and preferences reported by participants emphasizes the need for participatory design approaches to better tailor haptic technologies to users’ relational needs.

Advances in spatial computing give researchers new opportunities to engage participants in designing haptic technologies to support awareness. Augmented Reality (AR) headsets, like haptics, can provide awareness to users beyond the scope of a computer monitor. For example, a system similar to inPhase \cite{tsujita2010inphase} could transmit a flexible set of key gameplay events (e.g., loss of health, acquisition of resources, change in proximity) from one partner's game to another partner’s AR headset has potential to heighten awareness. Researchers could also utilize other basic haptic or AR devices in participatory design workshops with couples to generate awareness-promoting solutions. For example, couples could brainstorm how they would integrate haptics into specific joint game scenarios (e.g., how to represent their partner's presence during combat sequences, resource management, or exploration). 

\section{Limitations} A key limitation of this study is the relatively small number of couples interviewed, and correspondingly, the small number of couples that correspond to the developed archetypes. In addition, there may exist different variations of play dynamics that exist between couples but were not observed in the couples in this study, such as couples that may have a larger difference in their level of interest in playing games (e.g., one partner is more of a reluctant game player than the other), or couples that may interact with games in different ways, such as watching their partner play without playing themselves. Although we did not intend to recruit solely from an American context, our recruitment methods resulted in a primarily American context with few international participants. Different relationship norms exist in different cultures, so future work may explore cultural differences in relationships and gameplay. As we recruited for dyadic couples, these findings do not reflect the experience of people with non-normative relationship dynamics such as polyamory, which have been largely unexplored in HCI and design \cite{kinnee2024designing}.

\section{Conclusion and Future Work}

In this work, we investigated how 13 couples in long-distance relationships are using video games as a strategy for relational maintenance. We used a diary study, semi-structured interviews, and a mechanics-dynamics-aesthetics design activity to probe couples' values around how they use play to maintain emotional connection over distances often involving multiple time zones and thousands of miles. We observed differences in couple play styles and classified them into archetypes based on these variations. In addition, we build on a long legacy of HCI literature detailing how people appropriate technological affordances to express affection. Building on the design implications developed in this work, future work could involve exploring improved functionalities around facilitating memory making and physicalness in digital games for couples, including participatory design workshops for haptic experiences, co-design between game/platform developers and couples, or creating a system similar to the one we describe in Section \ref{memoriessection}. We contribute an empirical understanding of how couples in long-distance relationships are using games to maintain emotional connection, and provide design recommendations for game designers and researchers to create more technologies targeted at this growing population.

\begin{acks}
This work was supported by the University of Washington's Human Centered Design \& Engineering department. The authors would also like to thank the participants for sharing their experiences, and our directed research group for their valuable design insights. 
\end{acks}

\bibliographystyle{ACM-Reference-Format}
\bibliography{main}


\end{document}